\input{aipcheck}

\documentclass[
   ,final            
  ]
  {aipproc}

\layoutstyle{8x11double}

\usepackage{epsfig}\parskip 5pt plus 1pt
\usepackage{bbm}
\usepackage{amsmath}
\usepackage{amssymb}
\usepackage{amsfonts}
\usepackage{mathrsfs}
\usepackage{graphicx}
\newcommand{\raw}{\rightarrow}

\newcommand{\be}{\begin{equation}}
\newcommand{\ee}{\end{equation}}
\newcommand{\bea}{\begin{eqnarray}}
\newcommand{\eea}{\end{eqnarray}}
\begin{document}

\title{Non-unitary leptonic mixing and CP-violation}

 
\author{J. L\'opez-Pav\'on}{
  address={ Departamento de F\'\i sica Te\'orica and Instituto de F\'\i sica Te\'orica UAM/CSIC,
\\
Universidad Aut\'onoma de Madrid, 28049 Cantoblanco, Madrid, Spain}
}

\begin{abstract}
Some theories of new physics accounting for neutrino masses can give rise to a low-energy non-unitary leptonic mixing matrix. It is shown that the CP-asymmetries in the $\nu_\mu\rightarrow \nu_\tau$ channel are an excellent probe of such new physics. In adition, we clarify the relationship betweeen our framework and the so called "non-standard neutrino interactions" scenarios: the sensitivities explored here apply as well to such constructions, except for extremely fine-tuned cancellations.
\end{abstract}

\maketitle


\section{Introduction}
\label{intro}

\begin{small}
Non-zero neutrino masses are evidence for physics beyond the Standard Model (SM). The complete theory accounting for them and encompassing the Standard Model should be unitary, as mandated by probability conservation. The light fermionic fields in the theory may mix with other degrees of freedom. The complete mixing matrices are generically unitary. However, the effective $3 \times 3$ submatrices describing the mixing of the known fields need not to be unitary. 
We will assume that the full theory is indeed unitary, whereas, low-energy non-unitarity may result from BSM physics contributing to neutrino propagation, when the physical measurements are described solely in terms of SM fields~\cite{uni}. Specifically, the tree-level exchange of heavy fermions (scalars) will (not) induce low-energy non-unitary contributions~\cite{Abada:2007ux}.

  In Ref.~\cite{uni} the so-called MUV (minimal unitarity violation) was developed and the {\it absolute} values of the elements of the matrix N were determined. 
 However, no information on the size of the phases of the mixing matrix is available, neither on the standard ``unitary'' phases nor on the new non-unitary ones, as present oscillation data correspond mainly to disappearance experiments. It is the purpose of this work to explore the future sensitivity to the new CP-odd phases of the leptonic mixing matrix associated to the non-unitarity. Furthermore, a comparison between our approach and the results in the "non-standard neutrino interaction" scenarios~\cite{conchanir,yasuda,mascitas} will be performed. 
\end{small}

\section{Formalism}
\label{formalism}
\begin{small}Let us parameterize the general non-unitary matrix $N$, which relates flavor and mass fields, as the product of an hermitian and a unitary matrix, defined by
\be
\label{N}
\nu_{\alpha} = N_{\alpha i}\, \nu_{i}\equiv\left[ (1+\eta) U\right]_{\alpha i}\nu_{i}   \, ,
\ee
with $\eta^\dagger=\eta$. The bounds derived in Ref.~\cite{uni} for the modulus of the elements of $NN^\dagger$ also apply to the elements of $\eta$, since $NN^\dagger=(1+\eta)^2\approx 1+2\eta$ and it follows that
\medskip
\begin{eqnarray}
|\eta| =
\begin{pmatrix}
 <5.5\cdot 10^{-3}  &  < 3.5 \cdot 10^{-5}  &    < 8.0 \cdot 10^{-3} \\
 < 3.5 \cdot 10^{-5}  & <5.0\cdot 10^{-3}  &    < 5.1 \cdot 10^{-3} \\
 < 8.0 \cdot 10^{-3} &  < 5.1 \cdot 10^{-3}
&  <5.0\cdot 10^{-3}
\end{pmatrix},
\label{limits}
\end{eqnarray}
at the $90\%$ confidence level. The bound on $\eta_{\mu \tau}$ has been updated with the latest experimental bound on $\tau \raw \mu\gamma$~\cite{taumugamma}. Eq.~(\ref{limits}) shows that the matrix $N$ is constrained to be unitary, within a $10^{-2}$ accuracy or better. The unitary matrix $U$ in Eq.~(\ref{N}) can thus be identified with the usual unitary mixing matrix $U=U_{PMNS}$, within the same accuracy. The flavor
eigenstates can then be conveniently expressed as\footnote{The handy superscript $SM$ is an abuse of language, to describe the flavor eigenstates of the standard unitary analysis.}
\be
|\nu_\alpha> =
\dfrac{(1+\eta^*)_{\alpha \beta}}{\left[ 1+2\eta_{\alpha\alpha}+(\eta^2)_{\alpha\alpha}\right]^{1/2} }\,|\nu^{SM}_\beta> .
\label{estados}
\ee
It follows that the neutrino oscillation amplitude, neglecting terms quadratic in $\eta$, is given simply by
\bea
<\nu_\beta|\nu_\alpha (L)> =  A^{SM}_{\alpha \beta}(L)\,\left(1-\eta_{\alpha\alpha}-\eta_{\beta\beta} \right) +\nonumber\\
\sum_\gamma \left( \eta^*_{\alpha \gamma}A^{SM}_{\gamma \beta}(L)+
\eta_{\beta \gamma}A^{SM}_{\alpha \gamma}(L) \right)\,,
\label{amplitud}
\eea
\noindent
with
\be
A^{SM}_{\alpha \beta}(L)\equiv<\nu^{SM}_\beta|\nu^{SM}_\alpha (L)>
\ee
\noindent
being the usual oscillation amplitude of the unitary analysis.

New CP-violation signals arising from the new phases in $\eta$, require to consider appearance channels, $\alpha\ne\beta$.  The best sensitivities to such phases will be achieved in a regime where the first term in Eq.~(\ref{amplitud}) is suppressed. This happens at short enough baselines, where the standard appearance amplitudes become vanishingly small while $A^{SM}_{\alpha \alpha}(L)\simeq 1$. The total amplitude is then well approached by
\be
<\nu_\beta|\nu_\alpha (L)> =  A^{SM}_{\alpha \beta}(L)+ 2\eta_{\alpha \beta}^* + \mathcal{O}(\eta\,A),
\label{amp}
\ee
where $\mathcal{O}(\eta\,A)$ only includes appearance amplitudes and $\eta\,$ components with flavor indices other than $\alpha \beta$. Then, at short enough baselines, each oscillation probability in a given flavor channel, $P_{\alpha \beta}$, is mostly sensitive to the corresponding
$\eta_{\alpha \beta}$. The other elements of the $\eta$ matrix can be safely disregarded in the analyzes below, without implying to assume zero values for them. That is, their effect is generically subdominant, a fact that has been numerically checked for the main contributions. 

For instance, in a two family scenario and within the above-described approximation, the oscillation probability would read:
\be
P_{\alpha \beta} = \sin^{2}(2\theta)\sin^2\left(\frac{\Delta L}{2}\right)\nonumber
\ee
\be
-4|\eta_{\alpha \beta}|\sin\delta_{\alpha \beta}\sin(2\theta)\sin\left(\frac{\Delta L}{2}\right)
+4|\eta_{\alpha \beta}|^2,
\label{prob}
\ee
where $\Delta = \Delta m^2 /2E$ and $\eta_{\alpha \beta}=|\eta_{\alpha \beta}|e^{-i\delta_{\alpha \beta}}$.
The first term in Eq.~(\ref{prob}) is the usual oscillation probability when the mixing matrix is unitary. The third term is
the zero-distance effect stemming from the non-orthogonality of the flavor eigenstates. Finally, the second term is the CP-violating interference between the other two. Notice that, even in two families, there is CP-violation due to the non-unitarity.

In the section below, we will analyze the new sources of CP violation in the $\nu_\mu \rightarrow \nu_\tau$ channel, since present constraints on $\eta_{e \mu}$ are too strong to allow a signal in the $\nu_e \rightarrow \nu_\mu$ one (see Eq.~(\ref{limits})), and $\nu_e \rightarrow \nu_\tau$ has extra supressions by small standard parameters such as $\sin\theta_{13}$ or $\Delta_{12}$~\cite{paper}. When numerically computing $P_{\mu \tau}$, the only approximation performed will be to neglect all $\eta$ elements but $\eta_{\mu \tau}$. They should be indeed subdominant, as illustrated by Eq.~(\ref{amp}). In any case, we have checked this approximation numerically.

\section{Sensitivity to the new CP-odd phases: the $\nu_\mu \raw \nu_\tau$ channel}
\label{sensitivity}

As suggested by Eqs.~(\ref{prob}), the best sensitivities to  CP-violation  will be achieved at
short baselines and high energies, where the standard term is suppressed by $\sin^2(\frac{\Delta L}{2})$. We will therefore
study a Neutrino Factory beam~\cite{nf} resulting from the decay of $50$ GeV muons, to be detected at a $130$ Km baseline, which matches for
example the CERN-Frejus distance.  For these values, $\sin(\frac{\Delta_{31} L}{2})\simeq 1.7\cdot10^{-2}$ and $\sin(\frac{\Delta_{21} L}{2})\simeq 6\cdot10^{-4}$, where $\Delta_{jk}\equiv(m^2_j-m^2_k)/2E$.
All terms in the  oscillation probability Eq.~(\ref{prob}) can then be of similar order for the channel $\nu_\mu \raw \nu_\tau$, if the $\eta_{\mu\tau}$ value is close to their experimental limit in Eq.~(\ref{limits}). In what follows, we will assume $2\cdot10^{20}$ useful decays per year and five years running with each polarity, consider a 5 Kt Opera-like detector and, finally, sensitivities and backgrounds a factor 5 larger~\cite{paper} than those used for the $\nu_e\rightarrow\nu_\tau$ channel in Ref.~\cite{silver}.

In Ref~\cite{paper} the complete expanded expression for $P_{\mu\tau} $ can be found. However, as this probability is not suppressed by small standard parameters such as $\sin\theta_{13}$ or $\Delta_{12}$, the two family approximation in Eq.~(\ref{prob}), with $\theta=\theta_{23}$ and $\Delta=\Delta_{31}$, is very accurate to understand qualitatively the results.
That equation indicates that the CP-odd interference term is only suppressed linearly in $|\eta_{\mu \tau}|$.
\begin{figure}[t]
\vspace{-0.0cm}
\begin{tabular}{cc}
\hspace{-0.55cm} 
                 \includegraphics[width=5.4cm]{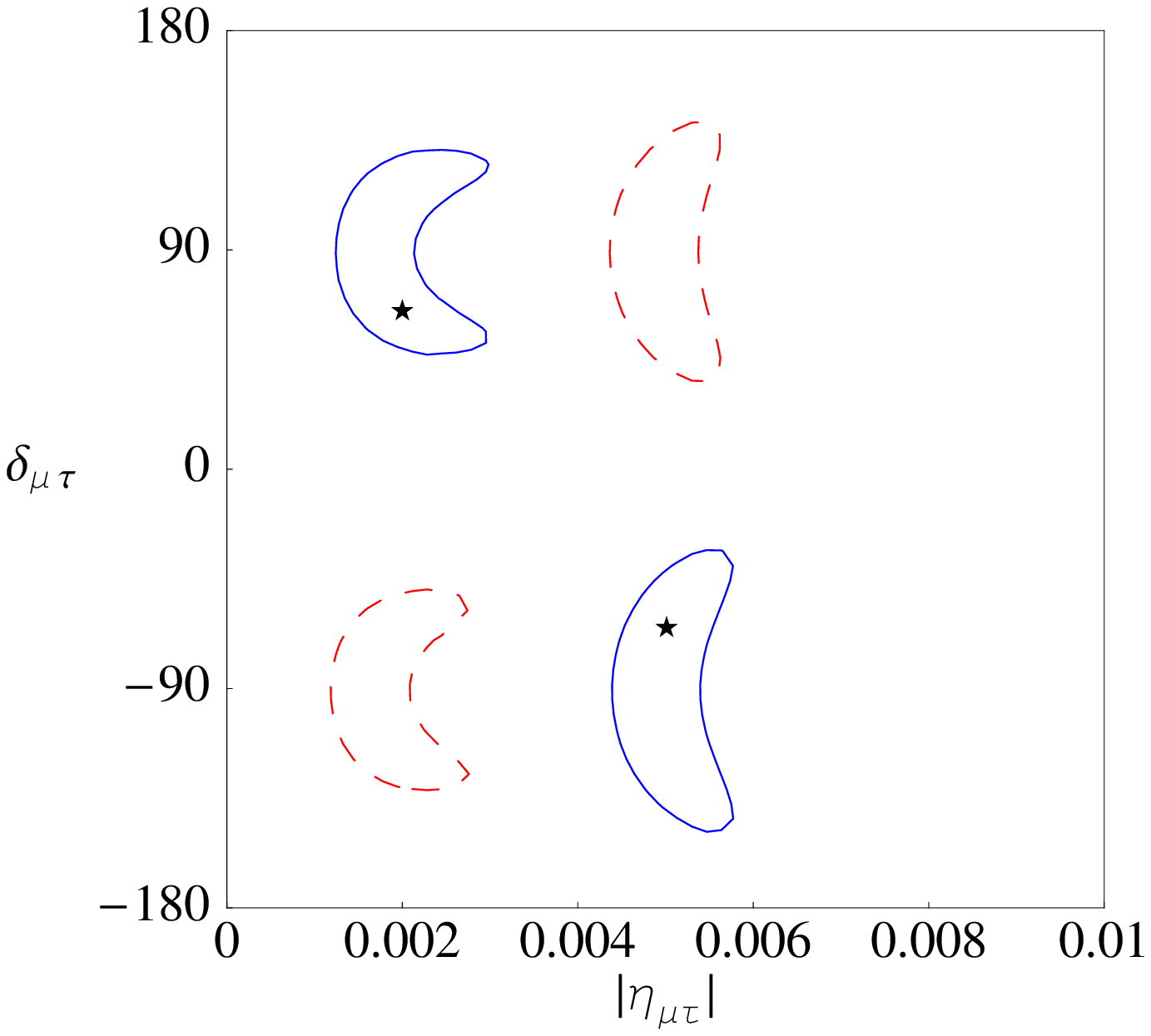} &
		 \includegraphics[width=5.4cm]{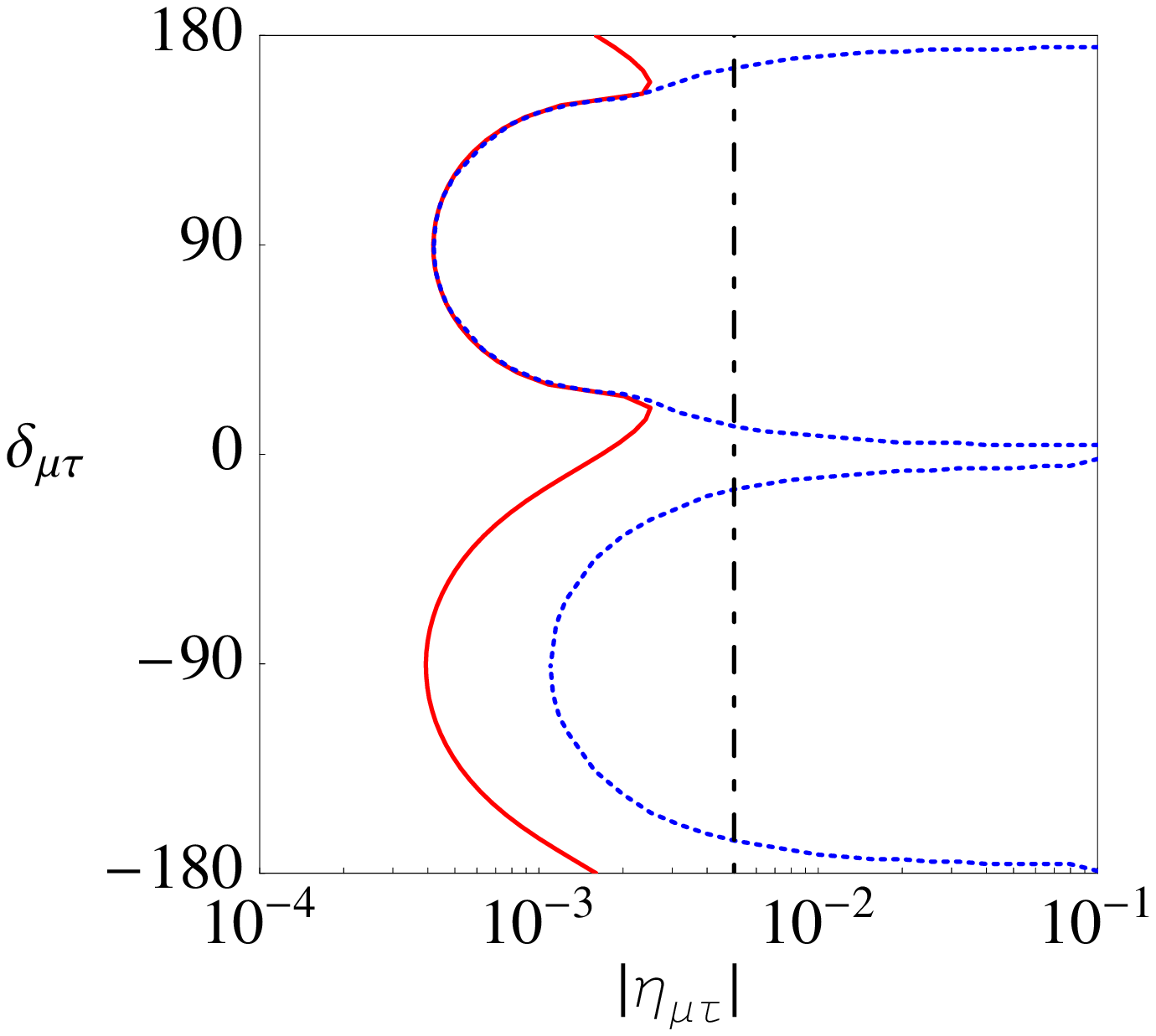}
\end{tabular}
\begin{tiny}
\caption{\it Left: $3 \sigma$ contours for two input values of $|\eta_{\mu \tau}|$ and $\delta_{\mu \tau}$ represented by
the stars. Right: the
solid line represents the $3\sigma$ sensitivity to $|\eta_{\mu \tau}|$ as a function of $\delta_{\mu \tau}$, the dotted
line the $3\sigma$ sensitivity to $\delta_{\mu \tau}$ and the dotted-dashed line represents
the present bound from $\tau \raw \mu \gamma$.
}\end{tiny}
\label{fig:mutau}
\end{figure}
This can indeed be observed in the result of the complete numerical computation, Fig.~\ref{fig:mutau}, which shows the sensitivities to $|\eta_{\mu \tau}|$ and $\delta_{\mu \tau}$ obtained. The left
panel represents two fits to two different input values of $|\eta_{\mu \tau}|$ and $\delta_{\mu \tau}$ (depicted by stars). The dashed lines correspond to fits done assuming the {\it wrong} hierarchy, that is the opposite sign for $\Delta_{31}$ to that with which the number of events were generated. As expected from Eq.~(\ref{prob}), a change of sign for the mass difference can be traded by a change of sign
for $\delta_{\mu \tau}$. Nevertheless, this does not spoil the potential for the discovery of CP violation, since a non-trivial value for  $|\delta_{\mu \tau}|$ is enough to indicate CP violation.
 Furthermore, the sinusoidal dependence implies as well a degeneracy between $\delta_{\mu \tau} \raw 180^\circ-\delta_{\mu \tau}$, as reflected  in  the figure.

The right panel in Fig.~\ref{fig:mutau} depicts the $3\sigma$ sensitivities to $|\eta_{\mu \tau}|$ (solid line) and $\delta_{\mu \tau}$ (dotted line), while the present bound from $\tau \raw \mu \gamma$ is also shown (dashed line). The poorest sensitivity to $|\eta_{\mu \tau}|$,
around $10^{-3}$, is found in the vicinity of $\delta_{\mu \tau} = 0$ and $\delta_{\mu \tau} = 180^\circ$, where the CP-odd interference term
 vanishes and the bound is placed through the subleading $|\eta_{\mu \tau}|^2$ term. The latter is also present at zero distance and its effects were already considered in Ref.~\cite{uni}, obtaining a bound of similar magnitude.
 The sensitivity to $|\eta_{\mu \tau}|$ peaks around $|\eta_{\mu \tau}| \simeq 4\cdot 10^{-4}$ for
$\delta_{\mu \tau} \simeq \pm 90^\circ$, where $\sin\delta_{\mu \tau}$ is maximum. That is, for non-trivial values of $\delta_{\mu \tau}$
not only CP-violation could be discovered, but values of $|\eta_{\mu \tau}|$ an order of magnitude smaller could be probed.
\end{small}
\section{Non-unitarity vs non-standard interactions}
\label{non-standard}
\begin{small}Non-standard neutrino interactions (NSI) are usually introduced through the addition of effective four-fermion operators to the SM Lagrangian~\cite{conchanir,yasuda,mascitas}, made out of the SM fields and invariant under the SM gauge group. They can affect the production or  detection processes or modify the matter effects in the propagation, depending on the operator or combination of operators considered. For instance, in Ref.~\cite{yasuda} new matter effects stemming from the operator\begin{small} $\left( \nu_{\alpha}^{SM} \gamma_{\mu} P_L\nu_{\beta}^{SM}\right)\left(\bar f\gamma_{\mu}f \right)$\end{small} were studied. In Ref.~\cite{conchanir}, another kind of four-fermion operators were introduced which affect the production and detection processes at a Neutrino Factory, but do not correct matter effects. In that case, defining $\nu_\alpha^{SM} \equiv U_{\alpha i}\,\nu_i$, with U being the PMNS matrix, the effective production and detection states are given by:
\begin{small}
\bea
|\nu_e^{p}> &=& (1+\epsilon^{*p})_{e\beta}|\nu^{SM}_\beta>=(1+\epsilon^{*p})_{e\beta}U_{\beta i}^* |\nu_i> \nonumber
\\
|\nu_\mu^d> &=& (1+\epsilon^{*d})_{\mu\beta}|\nu^{SM}_\beta>=(1+\epsilon^{*d})_{\mu\beta}U_{\beta i}^* |\nu_i>,\nonumber
\label{estadosconcha}
\eea
\end{small}
\noindent
where $\epsilon^{p(d)}_{\alpha\beta}$ are the coefficients of the new operators, up to normalization factors. These expressions are very similar to our parameterization of the effects of a non-unitary mixing matrix, Eq.~(\ref{estados}). In fact, the latter can also be encoded in terms of effective four-fermion operators, after integrating out the $W$ and $Z$ bosons. The difference is that, in the case of a non-unitary mixing matrix, the coefficients of the different effective operators induced by it and contributing to production, detection and matter effects are not independent but related. It follows from Eq.~(\ref{N}) that $\eta_{\alpha\beta}=\eta_{\beta\alpha}^*$. Which would mean $\epsilon_{\alpha\beta}^p = \epsilon_{\beta \alpha}^{d*}$, a constraint usually not required when introducing NSI. When such equality holds,  the oscillation physics induced by NSI is equivalent in vacuum to that stemming from  non-unitarity in the MUV scheme\footnote{In Ref.~\cite{uni} it was shown that matter effects are also modified in the presence of a non-unitarity mixing matrix, nevertheless, as we have studied a setup such that matter effects - standard and new ones - are negligible, the conclusion is the same even in matter.}. Furthermore, even if no such relations among  $\epsilon_{\alpha\beta}^p$ and $\epsilon_{\alpha \beta}^d$ are assumed, the order of magnitude of the bounds obtained in the previous chapters should apply as well to NSI, barring very fine tuned cancellations. The new CP signals analyzed in this work are, therefore, also probes of the phases of NSI.\end{small}

\section{Conclusions}
\label{conclusions}

\begin{small}
An asymmetry between the strength of $\nu_\mu\rightarrow \nu_\tau$ oscillations  versus that for $\bar\nu_\mu\rightarrow \bar \nu_\tau$ has been shown to be a beautiful  and  excellent probe of new physics, when measured at short-baselines ($\sim 100$ km.) using a Neutrino Factory  beam of energy $\mathcal{O}(20 GeV)$. Non-trivial values of the new phases can  lead not only to the discovery of
CP-violation associated to the new physics, but also allow to probe the absolute value of the moduli down to $10^{-4}$.

Our analyzes for future sensitivities to non-unitarity, as well as the new signals of CP-violation explored here,  also apply to the physics of ``non-standard neutrino interactions", except for extremely fine-tuned cancellations.
\end{small}





\bibliographystyle{aipproc}   

\begin{thebibliography}{9}
\bibitem{mascitas}A detailed list of papers about NSI can be found in Ref.~\cite{paper}
\end{thebibliography}


\IfFileExists{\jobname.bbl}{}
 {\typeout{}
  \typeout{******************************************}
  \typeout{** Please run "bibtex \jobname" to optain}
  \typeout{** the bibliography and then re-run LaTeX}
  \typeout{** twice to fix the references!}
  \typeout{******************************************}
  \typeout{}
 }

\end{document}